# Title: Weakly coordinating anions and solvents break the conductivity–transference–solvation tradeoff

Authors: *Alex Liu*[1†], *Ganesh Raghavendran*[1†], *Eric S. Yang*[2†], *Kevin J. Anderton*[2], *Vibhu Vardhan Singh*[1], *Aiden Larson*[1], *Kun Ryu*[3], *Kent J. Griffith*[2*], *Joshua S. Figueroa*[2*], *Ying Shirley Meng*[1,3*] *and Tod A. Pascal*[1*]

Affiliations:

[1]Aiiso Yufeng Li Family Department of Chemical and Nano Engineering, University of California San Diego, La Jolla, CA 92093, USA.

[2]Department of Chemistry, University of California San Diego, La Jolla, CA 92093, USA.

[3]Pritzker School of Molecular Engineering, University of Chicago, Chicago, IL 60637, USA.

**Corresponding author. Email:* tpascal@ucsd.edu (T.A.P.); shirleymeng@uchicago.edu (Y.S.M.); jsfig@ucsd.edu (J.S.F.); k3griffith@ucsd.edu (K.J.G.)

[†]These authors contributed equally to this work.

**Abstract:** Lithium-metal batteries promise high energy density, but common electrolytes trade ionic conductivity against the share of current carried by lithium ions and against how strongly solvent binds those ions. We show that simultaneously weakening both interactions of the lithium ion, with solvent and with anion, relaxes this trade-off. Thus, a weakly coordinating perfluorinated alkoxyaluminate salt in liquefied fluoromethane conducts 39.2 mS $cm^{-1}$ at 25 °C while maintaining a lithium-ion transference number of 0.71. Molecular dynamics simulations further reveal that the anion prevents persistent ion aggregation and lowers the salt's dissociation penalty. Consistent with these properties, nickel-rich cathode/lithium-metal anode full cells charge at 10C at room temperature and at 2C at −40 °C. We advance coordination strength as a design axis for future energy storage devices.

**Main Text:**

Lithium-ion batteries (LIBs) have transformed portable electronics, electric vehicles, and grid-scale systems (*1*), but their graphite anodes are approaching their theoretical capacity limit. Lithium-metal anodes offer a substantially higher specific capacity (3,860 mAh $g^{-1}$), but realizing this promise requires overcoming the intrinsic instability of lithium metal. Specifically, the highly reducing nature of lithium metal renders it incompatible with the conventional carbonate-based electrolytes used in LIBs (*2*), leading to uncontrolled dendrite growth and low Coulombic efficiency (*3*). This has motivated numerous mitigation strategies, including the application of three-dimensional host structures, artificial interfacial coatings, and electrolyte engineering (*4*–*6*). Among these approaches, electrolyte engineering is particularly promising.

Electrolyte performance is dictated by three interconnected transport properties: ionic conductivity (σ), $Li^+$ transference number ($t^+$), and $Li^+$ desolvation kinetics. Simultaneously optimizing these properties remains a longstanding challenge because the strategies that improve one parameter usually degrade the others. For example, increasing salt concentration adds charge carriers but also increases viscosity and promotes ion aggregation, reducing ion mobility and the $Li^+$ transference number (*7*, *8*). Conversely, weakly solvating solvents lower desolvation barriers but often promote contact-ion-pair formation (*9*–*11*).

Existing electrolyte designs illustrate this dilemma. Conventional carbonate-based electrolytes comprising lithium hexafluorophosphate ($LiPF_6$) and ethylene carbonate perform poorly at low temperatures because of their relatively high viscosity. High-concentration and localized high-concentration electrolytes (LHCEs) improve interfacial stability and safety (*12*, *13*), but at the cost of increased viscosity and reduced ionic conductivity (*14*, *15*). Liquefied-gas electrolytes offer ultralow viscosity, rapid ion transport, and inherent safety (*16–19*), but in these weakly solvating environments, the same reduction in solvent coordination that favors desolvation also promotes ion association.

Here, we overcome this trade-off by simultaneously weakening both the cation-solvent and the cation-anion interactions. We introduce a design paradigm that combines a highly dissociative weakly coordinating anion (WCA) salt with a weakly coordinating liquefied-gas solvent (WCS). Fluoromethane (FM) satisfies these solvent criteria, owing to its intrinsically weak coordinating ability and ultralow viscosity, resulting in weak $Li^+$ solvation and high $Li^+$ transport. By design, WCA salts possess large anionic radii, highly delocalized charge distributions, and low $Li^+$-binding affinities, enabling extensive salt dissociation. As a model system, we employ lithium tetrakis(perfluoro-tert-butoxy)aluminate (LiTPFA), whose weakly coordinating $[Al(OC(CF_3)_3)_4]^-$ anion remains highly dissociated even in low-dielectric media (*20*, *21*). The resulting WCA-WCS electrolyte outperforms benchmark formulations across key ion-transport metrics (*22*, *23*), achieving an ionic conductivity of 24.0 mS $cm^{-1}$ and a $Li^+$ transference number of 0.71, which in turn enables stable lithium-metal cycling, fast charging, and robust operation across a wide temperature range.

**Fast ion transport in fluoromethane-based WCA-WCS electrolytes**

We evaluate the factors governing ionic conductivity across a matrix comprising LiTPFA and four other conventional lithium salts [$LiPF_6$, lithium tetrafluoroborate ($LiBF_4$), lithium bis(fluorosulfonyl)imide (LiFSI) and lithium bis(trifluoromethanesulfonyl)imide (LiTFSI)] dissolved in FM and four other common organic solvents [acetonitrile (ACN), 1,2-dimethoxyethane (DME), diethyl ether (DEE) and dimethyl ether ($Me_2O$)]. The anions span a broad range of sizes (**Fig. S1**), charge delocalization (**Fig. S2**), and $Li^+$-binding affinities, and the solvents a wide range of dielectric constants and donor numbers (**Table S1**).

The traditional axis for enhancing ionic conductivity is the application of strongly coordinating high-dielectric solvents and low salt concentrations, resulting in solvent-separated ion-pairs (SSIP) with suppressed ion association (*24–26*). The WCA-WCS design paradigm (**Fig. 1A**) instead combines low ion correlation, high $Li^+$ transport, and weak $Li^+$ solvation. This choice has two defining consequences. In highly dissociating solvents such as ACN, LiTPFA offers no conductivity advantage. Here, smaller salts like LiFSI and $LiPF_6$ outperform it, reflecting the reduced mobility expected from its large size. We quantify this in **Fig. 1B**, where $k = \frac{r_-}{r_+}$ is the ratio of the volume-equivalent spherical radii of the isolated anion and of the $Li^+$–solvent cluster comprising its first solvation shell. Conductivity initially increases with anion size, consistent with the decrease of the dissociation free energy of the crystalline salt ($\Delta G_{diss}$), from $LiBF_4$ to $LiPF_6$ (**Fig. S3**), but decreases for the largest anions as reduced anion mobility and increased solution viscosity offset this benefit (*27*).

The advantage of WCA salts thus manifests most prominently in low-dielectric solvents such as FM and DEE (**Fig. 1B**), where weak cation-anion interactions limit ion association at low viscosity. The dielectric–fluidity factor ($\varepsilon/\eta$), which combines the two solvent properties, is a useful descriptor for identifying solvents that promote high charge diffusivity (**Fig. 1C**). Accordingly, 0.5 M LiTPFA/FM, with the highest calculated $\varepsilon/\eta$ of 120 $mPa^{-1}$ $s^{-1}$, reaches 39.2 mS $cm^{-1}$ at 25 °C and 23.4 mS $cm^{-1}$ at −60 °C (**Fig. 1D**). This performance exceeds state-of-the-art low-temperature electrolytes based on fluoroacetonitrile (*28*) and tetrahydropyran/1,2-dimethoxypropane mixtures (*29*), and surpasses the highest conductivity reported for the 1 M LiFSI/$Me_2O$ liquefied-gas electrolyte (*19*). At 25 °C, conductivity increases with salt concentration up to 0.5 M and then decreases up to the solubility limit of 1.4 M (**Figs. 1D** and **S4**). In subsequent studies at 0.25 M, where reduced ion association minimizes anion interference with $Li^+$ transport (*30*), LiTPFA/FM still exhibits a conductivity of 24.0 mS $cm^{-1}$ at 25 °C and 12.7 mS $cm^{-1}$ at −60 °C, the latter approaching the room-temperature conductivity of commercial electrolytes.

**High $Li^+$-selective transport in WCA-WCS electrolytes**

High-pressure liquefied-gas pulsed-field gradient-nuclear magnetic resonance (PFG-NMR) spectroscopy (**Fig. 2A**) quantifies self-diffusion across the electrolyte matrix and elucidates how salt and solvent properties govern $Li^+$ transport. In the ideal Nernst–Einstein limit (*31*), the $Li^+$ transport number ($t_{Li+}$) depends on the relative mobilities of the cation and anion. Here, according to the Stokes-Einstein relation, $t_{Li+}^{ideal} = \frac{(k)}{(k+1)}$ increases monotonically with $k$ and asymptotically approaches unity. Large, weakly coordinating anions are expected to favor $Li^+$ transport by reducing the relative contribution of anion motion to charge transport. Indeed, in strongly dissociating ACN, we measure $t_{Li+}$ values that closely track their ideal counterparts (**Fig. 2B**), consistent with near-ideal solution behavior and minimal ion association. Deviations from ideal behavior are expected to grow as solvent dielectric strength decreases, yet we find that LiTPFA deviates from its ideal transport number by less than 10% in every solvent tested (**Figs. 2B** and **S5**).

We parameterize a single-ion-pairing model (*32*) from the PFG-NMR diffusion coefficients to further quantify the parameters that affect $Li^+$ transport under an electric field. From the ionic self-diffusion and charge-diffusion coefficients, we calculate the inverse Haven ratio ($H^{-1}$), which quantifies how salt and solvent properties influence ion correlations (**Fig. 2C**). High-dielectric solvents such as ACN generally exhibit higher $H^{-1}$ values, consistent with weaker ion correlations, while low-dielectric solvents typically have $H^{-1}$ values that approach zero. Notably, LiTPFA maintains a high $H^{-1} \gtrsim 0.1$ across all solvents examined, indicating sustained, near ideal charge transport even in low-dielectric solvents. In contrast, $LiBF_4$ exhibits substantially lower $H^{-1} << 0.05$ in low-dielectric solvents, indicative of incomplete dissociation. LiTPFA is also unique in maintaining faster $Li^+$ vs. anion diffusion across every solvent investigated ($D_+/D_- = 1.08–1.24$) and is the only salt that retains appreciable ionicity in DEE (**Table S2**). In DEE we find that LiTFSI, LiFSI, and $LiPF_6$ depart from the consistency bounds of the single-ion-pairing model, attributable to the formation of higher-order ionic aggregates beyond the neutral 1:1 ion pairing assumed in the model.

Among all electrolytes studied, LiTPFA/FM exhibits the highest $t^+$ value, reaching 0.71 (**Fig. 2D**). While most electrolytes display transference numbers below their corresponding transport numbers, due to ion correlations and electrophoretic effects, LiTPFA-based electrolytes systematically approach or exceed their transport-number limits. We attribute this behavior to the highly delocalized charge distribution of $TPFA^-$, which weakens correlated ion motion and enhances the relative mobility of $Li^+$ under an applied electric field. Independently, direct measurement of the transference number using electrophoretic NMR (*33*), shown in **Fig. S6**, supports this interpretation, as does the observation that the transference number of LiTPFA/FM remained largely unchanged from -40 °C to 25 °C (**Fig S7**).

**Molecular mechanism and $Li^+$ solvation without ion aggregation**

We hypothesize that the WCA-WCS design leads to a comparatively low $Li^+$ desolvation penalty due to the weak solvent coordination and its interfacial structure, which we quantify computationally. Quantum-chemistry cluster calculations at 0 K (**Fig. 3A** and **B**) show that FM binds $Li^+$ most weakly of the five solvents: −26.1 kcal $mol^{-1}$ $molecule^{-1}$, against −37.0 to −52.6 kcal $mol^{-1}$ $molecule^{-1}$ for $Me_2O$, DEE, ACN, and DME (**Table S3**). Near room temperature, molecular dynamics simulations employing a polarizable fluctuating-charge force field (FQMD) reveal that in the bulk, $Li^+$ retains 4.1 ± 0.1 FM molecules in its first shell at 0.03–0.23 M, with no contact-ion-paired $TPFA^-$ (**Fig. S8**). The reversible work to lower the $Li^+$ coordination number by one contact equivalent in the bulk liquid at 293 K is $\Delta G_1$ = 2.6 $k_BT$ (~ 0.07 eV) in FM, lower than $\Delta G_1$ = 3.7 $k_BT$ in ACN. At a Li(100) electrode held at zero net charge, FM forms a dense first layer: its molecular-center density peaks 4.0 Å above the metal at about twice the bulk value, with a first minimum at 5.7 Å. The free-energy cost of a $Li^+$ ion reaching the outer edge of this first layer is only $\Delta G_{L1}$ = 2.9 ± 0.3 $k_BT$, with its coordination kept within 2 % of the bulk value, whereas bringing $Li^+$ into contact with the metal costs $\Delta G_C$ = 24.1 ± 0.3 $k_BT$ (**Fig. S9**). Together, these free energies suggest a WCA-WCS mechanism in which $Li^+$ presents as a solvent-separated, pre-reduction state next to the metal electrode. This would allow facile desolvation without incurring the large thermodynamic penalty of stripping strongly bound first-shell molecules that limit charge transfer in more conventional electrolytes (*34*).

Weak cation–anion interactions also suppress persistent ionic networks. FQMD simulations (**Fig. 3C** and **D**) show that LiTPFA in FM remains dispersed as 16 distinct ionic species throughout, with only transient ion contacts. In contrast, $LiPF_6$ rapidly aggregates into no more than two distinct ionic species among the 16 ions, and these aggregates persists for the entirety of the trajectory. Potential of mean force calculations of the ion-association free energy in FM (**Fig. 3E**) show that this distinct aggregation originates from fundamentally different ion-pairing thermodynamics. LiTPFA displays no thermodynamically stable contact-ion-pair state over the sampled separation range and a strongly unfavorable association free energy ($\Delta G_{assoc}$ = +15.0 ± 0.7 $k_BT$), whereas $LiPF_6$ exhibits a deep contact-ion-pair minimum and a strongly favorable association free energy ($\Delta G_{assoc}$ = −24.1 ± 0.4 $k_BT$). We find similar trends across the other four solvents (**Fig. S11** and **Table S4**). Given the observed insolubility of $LiPF_6$ and other conventional salts in FM (**Fig. S10**), we conclude that resistance to ion association is an intrinsic consequence of the weakly coordinating $TPFA^-$ anion rather than a unique feature of FM.

As a design principle, the free energy of mixing, $\Delta G_{mix}$, is a useful descriptor of electrolyte viability. Relative to the crystalline salt and neat solvent, $\Delta G_{mix}$ is the free-ion solvation free energy $\Delta G_{solv}$ plus the ion dissociation penalty $\Delta G_{diss}$, minus the ideal entropy of mixing $T\Delta S_{ideal}$ (**Eq. S6**, **Table S5**). Using this figure of merit, LiTPFA has the most favorable $\Delta G_{mix}$ of the salts computed and lies near the dissolution threshold in FM (**Fig. 3F**). Its advantage rests on $\Delta G_{diss}$, which is 29–46 $k_BT$ smaller for LiTPFA than for $LiPF_6$ and $LiBF_4$ because of the delocalized charge of $TPFA^-$, rather than on stronger solvation. In ACN, where all three salts dissociate and this component analysis can be checked against the ion-pairing free energies, LiTPFA retains the best $\Delta G_{mix}$ even though its relative free-ion solvation is $\Delta\Delta G_{solv}$ = 13 ± 4 $k_BT$ less favorable than that of $LiPF_6$. Across the five solvents, $\Delta G_{mix}$ ranks the three computed salts consistently with the measured conductivity or solubility. By contrast, the $Li^+$–solvent binding energy (**Table S3**) fails to predict salt viability, and the continuum solvation free energy (**Table S15**) predicts LiTPFA the least solvated salt in every solvent, at odds with its unique solubility in FM.

**Implications for electrochemical performance**

The transport and thermodynamic advantages of our WCA-WCS electrolyte translate into dramatic improvements in electrochemical performance. We demonstrate this experimentally with a 5 wt% $CO_2$ additive that stabilizes the lithium-metal interface (*16*, *18*, *22*). Owing to the extremely low dielectric constant of $CO_2$ ($\varepsilon r \approx 1.0$), we expect it to leave the solvation structure largely unchanged, and we observe a lithium-metal surface that remains pristine after prolonged immersion in the electrolyte (**Fig. S12**).

Li||Li symmetric cells show substantially lower polarization with the WCA-WCS electrolyte than with the ionic-liquid liquefied-gas electrolyte (IL-LGE) or localized high-concentration electrolyte (LHCE) benchmarks (**Fig. 4 A** and **B**). Averaged over 20 cycles at 8 mA $cm^{-2}$, the overpotentials are 164, 181, and 404 mV for WCA-WCS, IL-LGE, and LHCE, respectively. Impedance analysis from −78 °C to 40 °C is consistent with weak $Li^+$ solvation in the WCA-WCS. Namely, fitted to the equivalent circuit of **Fig. S13**, WCA-WCS shows lower ionic resistance of the electrolyte, $R_{electrolyte}$ (4.18 vs. 81.2 Ω for IL-LGE at -78 °C), and lower charge-transfer resistance at the Li-SEI interface, $R_{CT}$ (125 vs. $1.02 \times 10^3$ Ω under -78 °C). At -78 °C the impedance of $Li^+$ transport through the SEI, $W_{SEI}$, is higher for WCA-WCS at (247,000 vs. 67,500 Ω) and becomes comparable with increasing temperature (**Fig. S14**). The reduced polarization is consistent with the high conductivity, elevated $Li^+$ transference number, and weak $Li^+$ solvation we established above.

Improved transport characteristics of the WCA-WCS electrolyte are further reflected in the morphology of deposited lithium. Cryogenic focused-ion beam/scanning electron microscopy (cryo-FIB/SEM) after the final electrodeposition step at 8 mA $cm^{-2}$ reveals comparatively uniform Li deposition (**Fig. 4C**). We find dense and compact deposits for both WCA-WCS and IL-LGE, but highly dendritic growth in LHCE, consistent with its severe concentration polarization. In **Fig. S15**, we show that the interphase is carbonate-containing with LiF as a minor component, consistent with the $CO_2$-derived interphase reported for this solvent system (*22*). WCA-WCS thus promotes more homogeneous ion flux during deposition, which suppresses the localized depletion that commonly leads to unstable Li growth.

Critically, WCA-WCS enables stable LMB cycling over a wide temperature range. Cyclic voltammetry of Li||Al half cells demonstrate anodic currents and passivation behavior comparable with those of the conventional carbonate electrolyte LP57 up to 5 V vs. $Li/Li^+$ (**Fig. S16**). Li||NMC811 full cells demonstrate stable charging at rates from C/10 to 5C with a C/3 discharge (**Fig. 4D**). While all electrolytes deliver similar capacities at low rates, LiTPFA/FM retains 149 mAh $g^{-1}$ at 5C (approximately 13.6 mA $cm^{-2}$), 81% of the capacity under C/10 charging, essentially unchanged over 100 cycles. In contrast, both benchmark electrolytes suffer severe capacity losses at high rates. WCA-WCS also enables fast charging at 10C (≈ 118 mAh $g^{-1}$) near room temperature, where both benchmark electrolytes deliver negligible reversible capacity (**Fig. S17**). This advantage became even more pronounced at low temperature, where ion transport and desolvation kinetics typically become dominant rate-limiting factors (*34*). At a C/3 rate, Li||NMC811 full cells retained discharge capacities of 165, 115, and 78.4 mAh $g^{-1}$ at −20, −40, and −60 °C, respectively (**Fig. 4E**). At −40 °C, LiTPFA in FM supports 2C charging (approximately 6 mA $cm^{-2}$ at a nominal loading of ≈3 mAh $cm^{-2}$), delivering 96.6 mAh $g^{-1}$, 71%

of the 136 mAh $g^{-1}$ obtained at C/3 in the same rate series (**Fig. 4F**). Extended cycling at −40 °C and a 1C charge rate maintained approximately 112 mAh $g^{-1}$ after 50 cycles (**Fig. S18**). In comparison, previously reported ultra-low-temperature LMBs are commonly limited to C/10 or slower rates (*22*).

**Conclusions**

In summary, we establish a weakly coordinating anion--weakly coordinating solvent (WCA-WCS) electrolyte design paradigm that resolves longstanding trade-offs among ionic conductivity, $Li^+$ transference, and $Li^+$ solvation in liquid electrolytes. By pairing a highly dissociative weakly coordinating salt (LiTPFA) with an ultra-low-viscosity liquefied-gas solvent (FM), we access a transport and electrochemical performance regime that has not been achieved using conventional electrolyte design strategies. The two components play complementary roles: weak solvent coordination keeps $Li^+$ solvation weak, while the charge-delocalized anion suppresses persistent ion association and reduces the thermodynamic cost of salt dissolution, so that the electrolyte avoids aggregation despite its low dielectric environment. Relative to state-of-the-art electrolytes, the WCA-WCS electrolyte already approaches and, in some cases, exceeds key performance metrics established for future energy storage devices. It enables low polarization during lithium-metal cycling, uniform lithium deposition for enhanced safety, and charging rates as high as 10C at room temperature and 2C at −40 °C, conditions where ion transport and interfacial charge transfer typically become severe limitations.

More broadly, these results identify weakly coordinating ion-solvent chemistries as a general route for decoupling electrolyte properties that have historically been constrained. Because the underlying design principles are not specific to lithium chemistry, the WCA-WCS framework may provide a foundation for electrolytes for other metal-anode systems, including sodium-, potassium-, and zinc-based batteries. Beyond batteries, this design principle should apply equally to any electrochemical energy-storage technology whose performance is limited by ion–solvent and ion–anion association, including electric double-layer capacitors and redox-flow cells. While the perfluorinated nature of LiTPFA serves as an ideal model system, this framework now provides the theoretical blueprint for designing next-generation, halogen-free WCAs that meet emerging environmental and scalability targets.

Materials and methods are available in supplementary materials.

**Acknowledgments:** Experimental portions of this work were performed in part at the San Diego Nanotechnology Infrastructure (SDNI) of UC San Diego, which is supported by the National Science Foundation (NSF), Grant ECCS-1542148. The authors acknowledge the use of facilities and instrumentation supported by NSF through the UC San Diego Materials Research Science and Engineering Center (UCSD MRSEC), DMR-2614051. This research used resources of the National Energy Research Scientific Computing Center, a DOE Office of Science User Facility supported by the Office of Science of the U.S. Department of Energy under Contract No. DE-AC02-05CH11231 using NERSC awards BES-ERCAP0037331 and NERSC DDR-ERCAP0039121. This work also used the Expanse supercomputer at the San Diego Supercomputer Center through allocations TG-NNT230002, TG-MAT240064, TG-MAT250088 and TG-NNT240007 [T.A.P.] and MAT250051 and MAT250127 [Y.S.M.] from the Advanced Cyberinfrastructure Coordination Ecosystem: Services & Support (ACCESS) program, which is supported by U.S. National Science Foundation grants #2138259, #2138286, #2138307, #2137603, and #2138296.

*Use of AI-assisted technologies.* Large language model tools were used as aids in this work and are disclosed here in accordance with Science’s policy. Claude (v2.1.299, Anthropic; through Claude Code) was used by V.V.S. to write and manage simulation workflows and to perform independent internal consistency checks of results against the underlying data. Codex and ChatGPT (GPT 6.0 Astra model, OpenAI) were used by G.R. and A.Li. to write and run the code that computes the van der Waals volumes and effective ionic radii, and to run the peak fitting of the XPS spectra. No AI tool was used as a physical model: all simulations used the force fields and electronic-structure methods described in the supplementary materials. All AI-assisted text, code and analyses were reviewed by the authors, who verified every reported number against the underlying data and take full responsibility for the content.

**Funding:** This work was primarily supported by the Energy Storage Research Alliance (ESRA), DE-AC02-06CH11357 and discretion fund by Sustainable Power and Energy Center (SPEC) at UC San Diego. Y.S.M. acknowledges partial support from startup funding at the University of Chicago. T.A.P. acknowledges partial support from the Alfred P. Sloan Foundation.

**Author contributions:** Conceptualization: G.R., Y.S.M., J.S.F., K.J.G., T.A.P. Methodology: G.R., E.S.Y., J.S.F., Y.S.M. Investigation: G.R., A.Li., A.La., K.J.A., V.V.S., K.R. Formal analysis: K.J.A., G.R., A.Li. Software: V.V.S. Visualization: G.R., A.Li., V.V.S. Supervision: Y.S.M., J.S.F., K.J.G., T.A.P. Funding acquisition: Y.S.M., J.S.F., K.J.G., T.A.P. Writing – original draft: G.R., A.Li., T.A.P. Writing – review & editing: all authors.

**Competing interests:** G.R., E.S.Y., J.S.F., Y.S.M. and T.A.P. are inventors on US provisional patent application no. 63/782,767, filed by the Regents of the University of California and covering the WCA-WCS electrolyte described here. The other authors declare that they have no competing interests.

**Data, code, and materials availability:** Tabulated data underlying all figures are provided as data S1. The data, scripts, simulation inputs and force-field parameters that reproduce the figures are deposited at Zenodo (https://doi.org/10.5281/zenodo.23676154) (*35*). Materials are described in the Materials and Methods; no new materials were generated beyond those described there.

**List of Supplementary Materials:** Materials and Methods; Supplementary Text; Figs. S1 to S20; Tables S1 to S16; References (36–66); Data S1

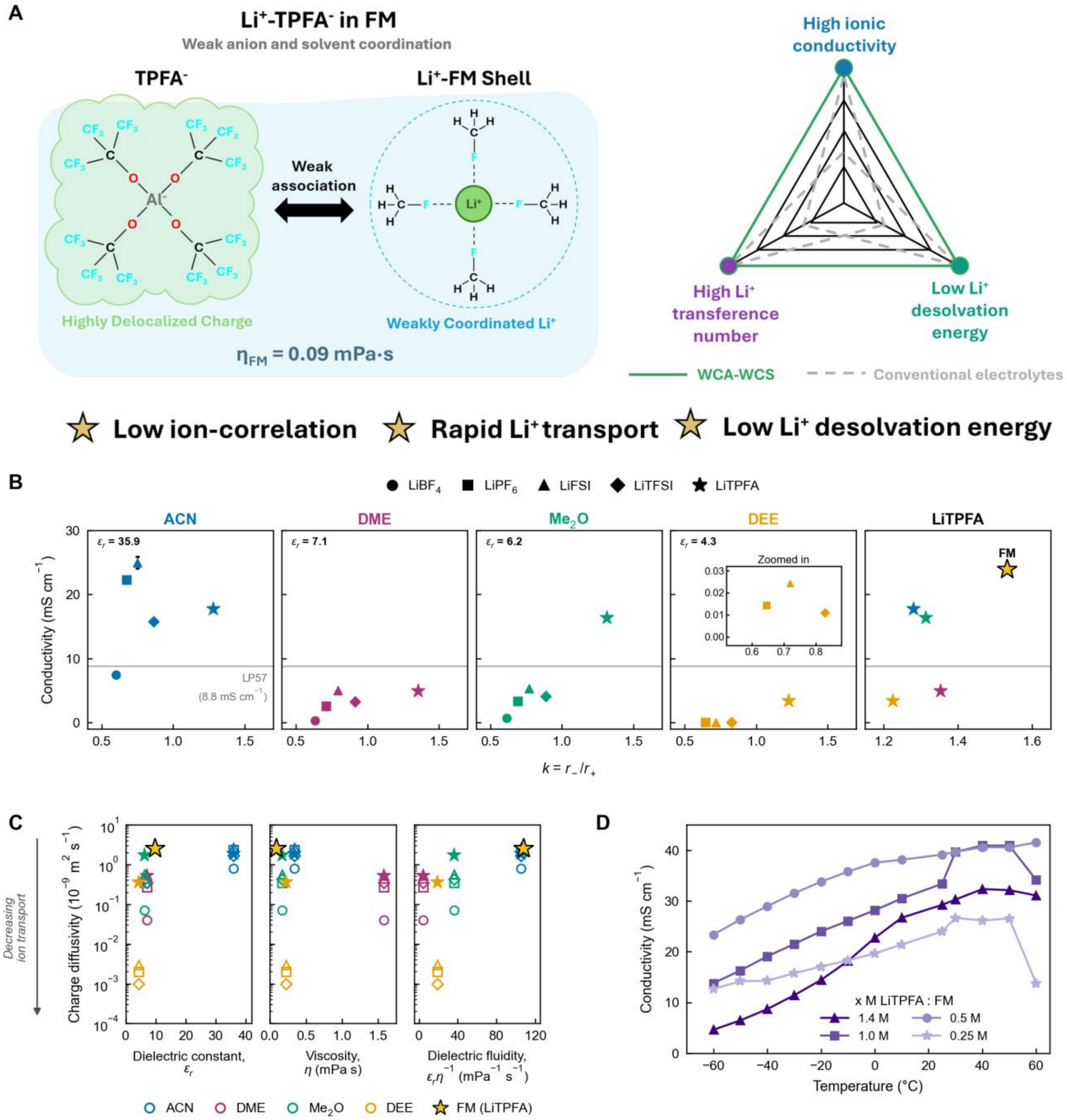

**Fig. 1. Weakly coordinating anion–weakly coordinating solvent framework and performance.** **(A)** Design schematic: weak anion and solvent coordination give high ionic conductivity, high $Li^+$ transference number and low $Li^+$ desolvation energy without the typical design trade-off. **(B)** Ionic conductivity of 0.25 M solutions of five lithium salts in ACN, DME, DEE, $Me_2O$ and FM at 25 °C against the anion-to-cation radius ratio $k = r_-/r_+$; error bars are the reported uncertainties. **(C)** Dependence of charge diffusivity of 0.25 M solutions on dielectric constant, viscosity and dielectric-fluidity factor. **(D)** Temperature-dependent conductivity of LiTPFA/FM at 0.25, 0.5, 1.0 and 1.4 M (−60 °C to 60 °C).

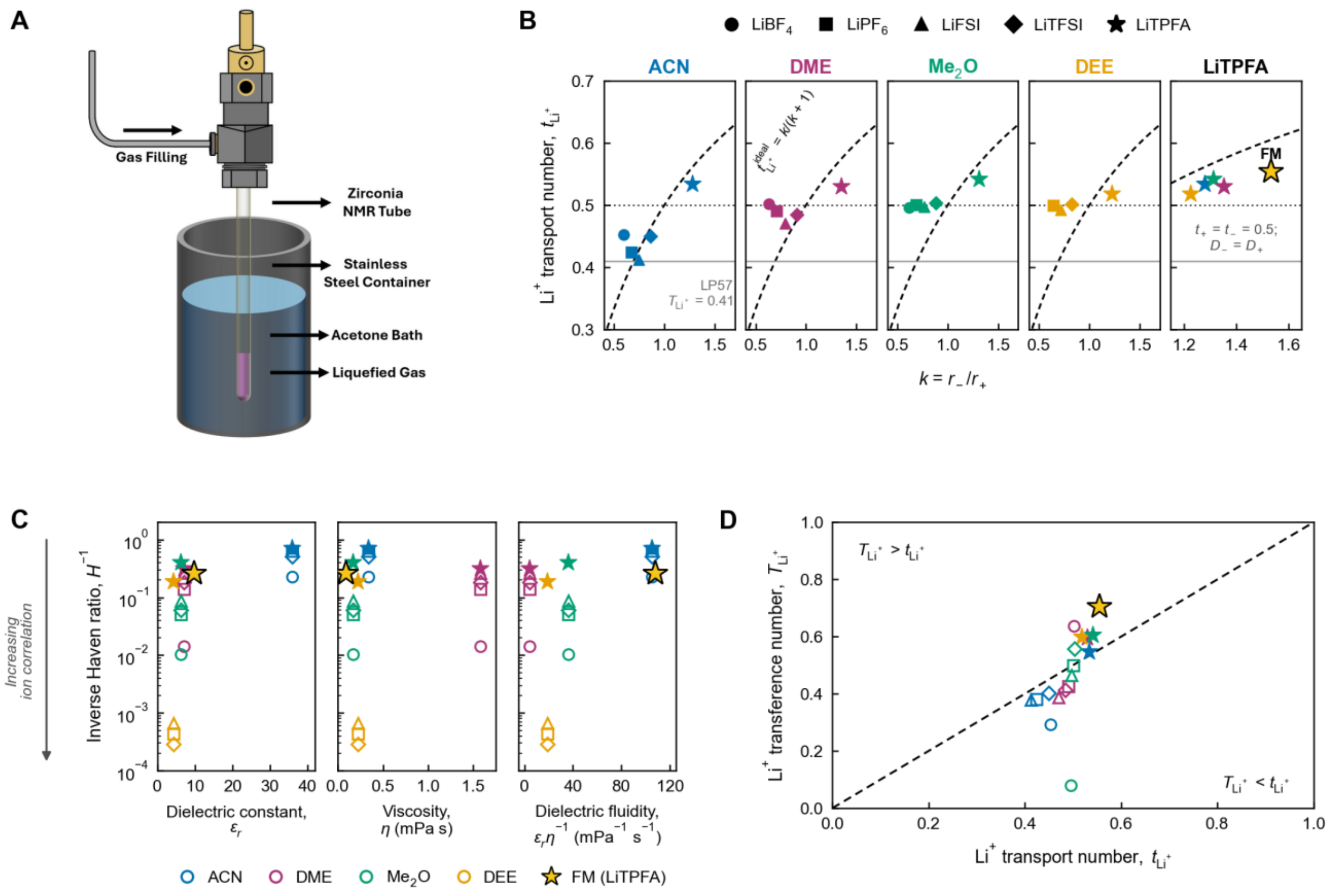

**Fig. 2. Ion transport in 0.25 M electrolytes at** 298 **K. (A)** Schematic of the liquefied-gas PFG-NMR setup: the gas is admitted through the gas-filling line and valve into a high-pressure zirconia NMR tube, immersed in an acetone bath (cooled with dry ice to −78 °C) inside a stainless-steel container. **(B)** $Li^+$ transport number against the anion-to-cation radius ratio k = $r_–/r_+$ for 0.25 M solutions of five lithium salts in ACN, DME, DEE, $Me_2O$ and FM at 25 °C; dashed curves, the ideal relation k/(k + 1). **(C)** Dependence of inverse Haven ratio of 0.25 M solutions on dielectric constant, viscosity and dielectric-fluidity factor. **(D)** Transference number against transport number, with the $y = x$ diagonal shown as a dashed line; points above the line have transference exceeding transport. $LiPF_6$, LiFSI and LiTFSI in DEE, for which the model returns a transference number outside 0 to 1, are omitted (table S2). Color denotes solvent and shape denotes salt throughout, matching Fig. 1B.

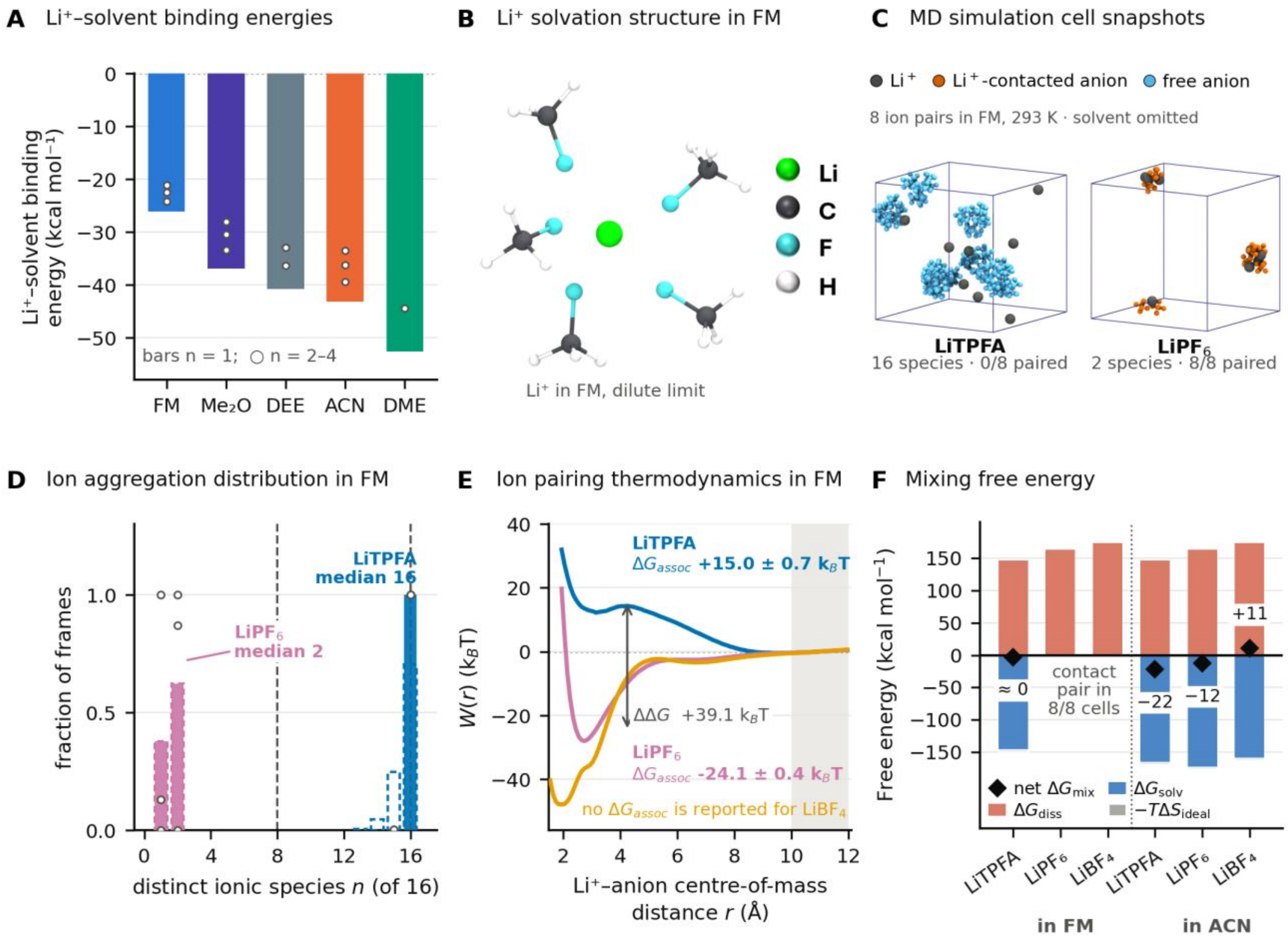

**Fig. 3. WCA-WCS microscopic thermodynamics. (A)** $Li^+$–solvent binding energies from gas-phase cluster calculations (bars, one solvent molecule; circles, two to four). **(B)** Illustrative $Li^+$ solvation structure in FM, dilute limit. **(C)** Simulation cells for LiTPFA and $LiPF_6$ in FM at 293 K, eight ion pairs each, solvent omitted; anions colored by contact with a $Li^+$. Each cell is the frame nearest its median species count. **(D)** Distribution of the number of distinct ionic species among the sixteen ions, for LiTPFA and $LiPF_6$; dashed rules mark all ions paired (n = 8) and no contact (n = 16). Bars, 3.0 Å contact cut-off; outlines, 4.5 Å; circles, the three replicas. **(E)** $Li^+$–anion potentials of mean force W(r) in FM at 293 K, zeroed on 10–12 Å (shaded); bands, ±1 s.d. over eight independent biases; $\Delta G_{assoc}$ evaluated at a uniform bound-state radius of 4.25 Å. $LiBF_4$ $\Delta G_{assoc}$ (table S4) is not shown (different lower wall and, in FM, anion–solvent parameterization; SM section I.1). **(F)** Mixing free energy $\Delta G_{mix}$ at 293 K: dissociation penalty $\Delta G_{diss}$, solvation free energy of the simulation cell $\Delta G_{solv}$ ($Li^+$–anion energy included) and $-T\Delta S_{ideal}$; diamonds, net $\Delta G_{mix}$. $LiPF_6$ and $LiBF_4$ contact-pair in every FM cell: only $\Delta G_{diss}$ shown.

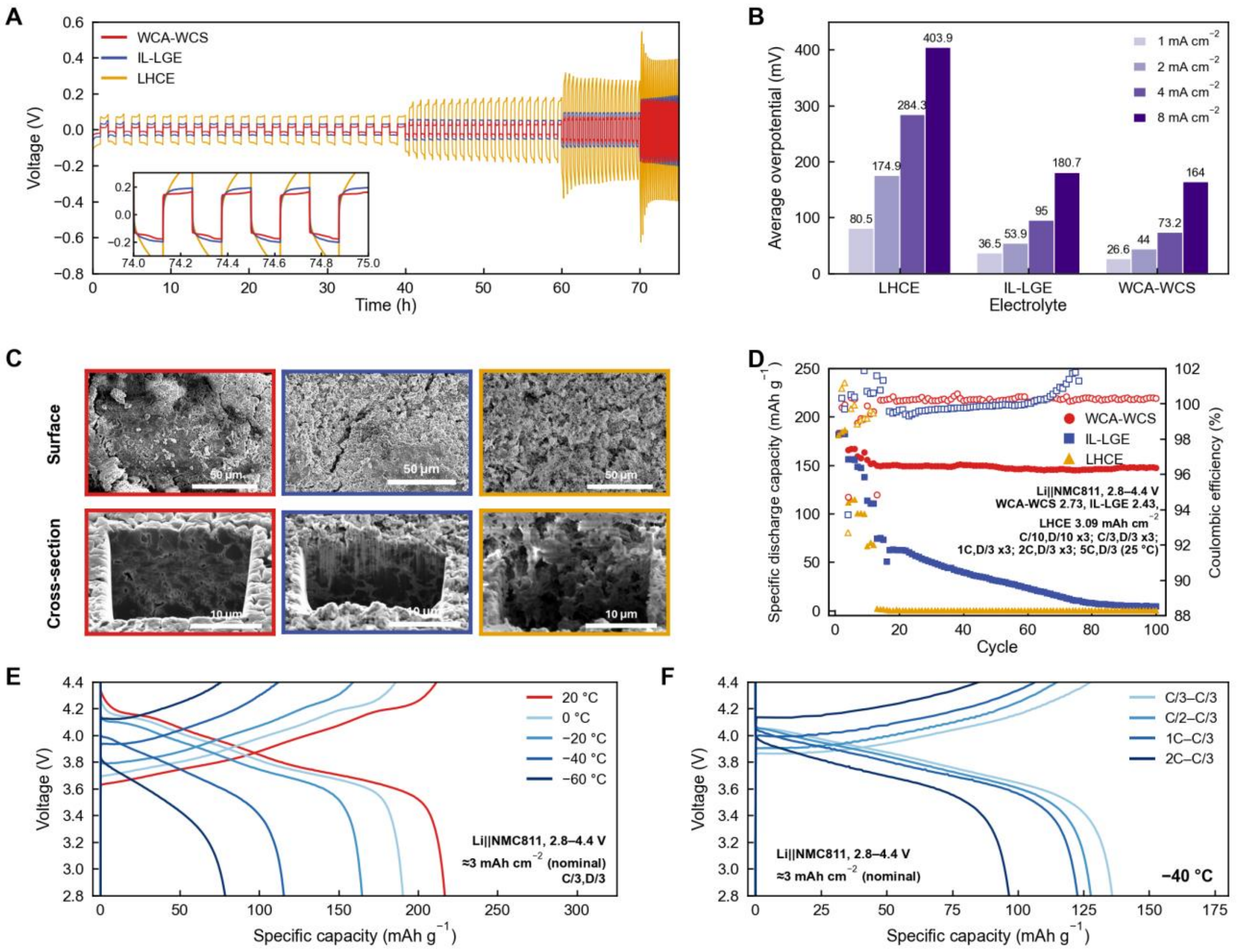

**Fig. 4. Electrochemical performance of the WCA-WCS electrolyte.** **(A)** Li‖Li symmetric cells in WCA-WCS (red), IL-LGE (blue) and LHCE (orange) at 1, 2, 4 and 8 mA $cm^{-2}$, 1 mAh $cm^{-2}$ per half-cycle and 25 °C; the inset expands the final hour at 8 mA $cm^{-2}$. **(B)** Average overpotential over 20 cycles for each electrolyte, colored by current density. **(C)** Cryo-FIB/SEM of the plated Li after the final deposition at 8 mA $cm^{-2}$, for WCA-WCS (red), IL-LGE (blue) and LHCE (orange). **(D)** Li‖NMC811 cells in WCA-WCS (red circles), IL-LGE (blue squares) and LHCE (orange triangles) charged at rates from C/10 to 5C against a fixed C/3 discharge, at 25 °C; filled symbols are discharge capacity, open symbols the Coulombic efficiency on the right axis. **(E)** Li‖NMC811 in WCA-WCS charged and discharged at C/3 from 20 to −60 °C, and **(F)** a rate series at −40 °C, charged at C/3 to 2C against a C/3 discharge (C/3 capacity 136 mAh $g^{-1}$, against 115 in E). WCA-WCS is LiTPFA/FM with a 5 wt% $CO_2$ additive; IL-LGE is the ionic-liquid liquefied-gas electrolyte and LHCE the localized high-concentration electrolyte.